# THE IMPACT OF MOBILE NODES ARRIVAL PATTERNS IN MANETS USING POISSON MODELS

John Tengviel[1], K. A. Dotche[2] and K. Diawuo[3]

[1]Department of Computer Engineering, Sunyani Polytechnic, Sunyani, Ghana
john2001gh@yahoo.com
[2]Department of Telecommunications Engineering, KNUST, Kumasi, Ghana
kdotche2004@gmail.com
[3]Department of Computer Engineering, KNUST, Kumasi, Ghana
kdiawuo@yahoo.com

*Abstract*

*Mobile Ad hoc Networks (MANETs) are dynamic networks populated by mobile stations, or mobile nodes (MNs). Specifically, MANETs consist of a collection of nodes randomly placed in a line (not necessarily straight). MANETs do appear in many real-world network applications such as a vehicular MANETs built along a highway in a city environment or people in a particular location. MNs in MANETs are usually laptops, PDAs or mobile phones. These devices may use Bluetooth and/or IEEE 802.11 (Wi-Fi) network interfaces and communicate in a decentralized manner. Mobility is a key feature of MANETs. Each node may work as a router and the network can dynamically change with time; when new nodes can join, and other nodes can leave the network.*

*This paper presents comparative results that have been carried out via Matlab software simulation. The study investigates the impact of mobile nodes' parameters such as the speed, the arrival rate and the size of mobile nodes in a given area using Poisson distribution. The results have indicated that mobile nodes' arrival rates may have influence on MNs population (as a larger number) in a location.*

## 1. INTRODUCTION

Mobile Ad-hoc NETworks (*MANETs)* is a collection of wireless mobile nodes configured to communicate amongst each other without the aid of an existing infrastructure. MANETS are *Multi-Hop* wireless networks since one node may not be indirect communication range of other node. In such cases the data from the original sender has to travel a number of hops (hop is one communication link) in order to reach the destination. The intermediate nodes act as routers and forward the data packets till the destination is reached [1, 14].

Like other networks the performance of ad hoc networks is affected by its topology. However, in ad-hoc networks the role of topology becomes more critical due to nodes' mobility [1, 2].

Consequently, many simulation tools are being used for ad hoc networks studies with help of mathematical models known as mobility models to generate various kinds of network topologies. The initial entry point and distribution of a *Mobile Node (MN)* may be far random and may be influenced by various factors (time of day, traffic conditions, weather conditions, rescue mission, and so on).

Ad hoc networks are viewed to be suitable for all situations in which a temporary communication is desired. The technology was initially developed keeping in mind the military applications [3] such as battle field in an unknown territory where an infrastructured network is almost impossible to have or maintain. In such situations, the ad hoc networks having self-organizing [4, 14] capability can be effectively used where other technologies either fail or cannot be effectively deployed. The entire network is mobile, and the individual terminals are allowed to move freely. Since, the nodes are mobile; the network topology is thus dynamic. This leads to frequent and unpredictable connectivity changes. In this dynamic topology, some pairs of terminals may not be able to communicate directly with each other and have to rely on some other terminals so that the messages are been delivered to their destinations. Such networks are often referred to as multi-hops or store-and-forward networks [5, 14].

This paper presents a study on mobile nodes in MANETs using Poisson models. We have decided to use the Poisson distribution since it has some properties that make it very tractable to mobile nodes arrival pattern. Though not very realistic from a practical point of view, a model based on the exponential distribution can be of great importance to provide an insight into the mobile nodes arrival pattern. The section 2 illustrates a brief review on MANETs studies. The section 3 introduces the Poisson distribution models. The simulation procedures and considered parameters are presented in section 4. The obtained results are objects in section 5 and the section 6 conclude the paper to further research works.

## 2. RELATED WORKS

The study of mobility models have shown that currently there are two types used in simulation of networks [6, 7]. These are traces and synthetic models. Traces are those mobility patterns that are observed in real-life systems. Traces provide accurate information, especially when they involve a large number of mobile nodes (MNs) and appropriate long observation period. On the other hand, synthetic models attempt to realistically represent the behaviour of MNs without the use of traces. They are divided into two categories, entity mobility models and group mobility models [8, 9]. The entity mobility models randomise the movements of each individual node and represent MNs whose movements are independent of each other. However, the group mobility models are a set of groups' nodes that stay close to each other and then randomise the movements of the group and represent MNs whose movements are dependent on each other. The node positions may also vary randomly around the group reference point. In [10], the mobility study in ad hoc has been approximated to pedestrian in the street, willing to exchange content (multimedia files, mp3, etc.) with their handset whilst walking at a relative low speed. All pedestrians have been assumed to be within a predefine range of communication and do not collide with each other as it is a case in a dense network. The semi-analytic study has shown that it is possible for communication devices in urban areas to be made up of an efficient MANETs where they work and efficiently share content through a unique server. It is important to recall that Mobility models should be able to mimic the distribution and movement pattern of nodes as in real-life scenarios. Some researchers have proposed indigenously mobility models such as Random Walk,

Random Waypoint, [3, 4], etc. for performance comparison of various routing protocols. The concern with indigenously designed models is that they represent a specific scenarios not often found in real lives. Hence their use in ad hoc network studies is very limited. Random Walk or Random Waypoint model though simple and elegant, produce random source of entry into a location with scattered pattern around the simulation area. In real-life, this may not really be the case.

## 3. MODELS OF STUDY

### 3.1. POISSON ARRIVAL DISTRIBUTION (NUMBER OF NODES)

When arrivals occur at random, the information of interest is the probability of *n* arrivals in a given time period, where *n* = 0, 1, 2, . ……n-1
Let $\lambda$ be a constant representing the average rate of arrival of nodes and consider a small time interval $\Delta t$, with $\Delta t \to 0$. The assumptions for this process are as follows:

- The probability of one arrival in an interval of $\Delta t$ seconds, say **(t, t+$\Delta$t)** is $\lambda \Delta t$, independent of arrivals in any time interval not overlapping **(t, t+$\Delta$t).**
- The probability of no arrivals in $\Delta t$ seconds is **1-$\lambda\Delta$t**, under such conditions, it can be shown that the probability of exactly **n** nodes arriving during an interval of length of **t** is given by the Poisson distribution law [12-14] in equation 1:

$$P(n) = \frac{(\lambda t)^n e^{-\lambda t}}{n!}, \quad \text{where } n \geq 0, t > 0 \qquad 1$$

The assumption of Poisson MN arrivals also implies a distribution of the time intervals between the arrivals of successive MN in a location. Let $\tau$ being greater than or equal to the time interval *t*, so from Equation 1:

$$P(0) = P[\tau > t] = e^{-\lambda t} \qquad 2$$

This MNs distribution in a location is known as the negative exponential distribution and is often simply referred to as the exponential distribution.

### A. INTER-ARRIVAL TIME EXPONENTIAL DISTRIBUTION

This method of arrival specifies the time between arrivals. In this case one indicates the probability distribution of a continuous random variable which measures the time from one arrival to the next. If the arrivals follow a Poisson distribution, it can be shown mathematically that the interarrival time will be distributed according to the exponential distribution [13, 14].
Let assume, if $P[t_n>t]$ is just the probability that no arrivals occur in (0, t), that is $P_0(t)$. Therefore, we have $A(t) = 1 - P_0(t)$. That is the Probability Distribution Function [PDF] is given as in equation 3.

$$A(t) = 1 - e^{-\lambda t} \text{ Or } PDF[A(t)] = 1 - e^{-\lambda t}, t \geq 0 \qquad 3$$

and its probability density function (pdf) is given by the equation 4.

$$f(t) = \frac{\partial A(t)}{\partial t} = \lambda e^{-\lambda t}, \ t \geq 0 \qquad\qquad 4$$

## 4. METHODOLOGY

### 4.1. Queueing System

The queueing system can be described as in Figure 1. by the following characteristics:

   i. Arriving Nodes: this is specified by the distribution of inter-arrivals time of nodes, that is arrival patterns
   ii. Exiting Nodes: it is specified by the distribution of the time taken to complete service, which is assume to be departures
   iii. Server: It has a single server or location
   iv. Input Source: The number of nodes that arrive in the service facility
   v. Queueing Discipline: The first – comes – first serves (FCFS) is assumed as the service discipline.

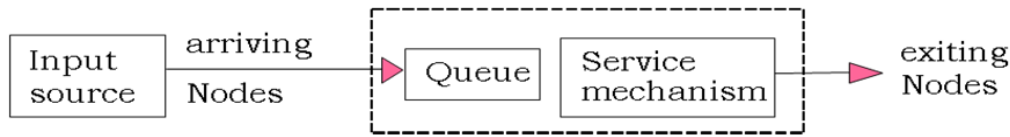

Figure 1: M/M/1 Queueing System Model

The Queueing Theory Model allows probabilistic predictions of the MNs movement. The system model can also be illustrated as in Figure 1. The Assumptions on the mobile nodes and the network in this study are:

- Independent and identically distributed (IID) - Stationary (Time homogeneity),
- The system (location) consists of a set of n independent MNs communicating over a wireless network.
- All communication links are bidirectional, i.e., all nodes have the same transmission range,
- Nodes communicate without interference and collision, i.e., a free space propagation model is considered.

### 4.2. Simulation Tools and Considerations

The section deals with the simulations in order to study the performance indicators of the queueing mobility model. A brief description of the simulation environment, the metrices collected and the various simulations perform follows. The computer program (or simulation) was run with several random values and the modeled behaviours were recorded for analysis and displayed in form of snapshot.

### 4.2.1. MATrix LABoratory (MATLAB) Simulator

The simulations have been carried out by using MatLab [15]. Matlab is developed by MathsWorks. It is commercially available. It is a high performance language for technical computing mostly used by engineers and researchers. It can be integrated to oriented languages as C++, and contains a variety of tools capable of computation and mathematical modeling, algorithm development and analysis, modeling and simulation, data analysis etc. Hardware and operating system (OS) configurations for performing simulations were specified as in Table 1

Table 1: Hardware and Software Configuration

| Processor | Pentium D, CPU 3GHz |
|---|---|
| RAM | 8GB |
| Hard Drive | 350GB |
| Operating System (OS) | Windows Vista/7 |
| MatLab | 8 |

### 4.2.2. Algorithm

The majority of current research in MANETs assumes that the nodes are uniformly randomly placed over the system area. Such a homogeneous initial point of entry is convenient in the network simulations. In real networks, however, the nodes are in general not uniformly distributed initially but have a common source of entry or exit. We therefore model the nodes initial entry point by designing inhomogeneous node placement for both QMM and RWMM. We also considered the important properties of the resulting distribution as the simulation progresses, as well as the probability density of the arrival patterns of the MNs. The section also considered the algorithms and flowcharts for the simulation presented. The QMM uses arrival and departure rates to calculate the queue mobility part from time t to time t+1, which also include the RM component. The new position for each MN is then calculated by summing the random motion with the new queueing mobility. Below are the algorithms and flowcharts for the simulation of both QMM and Random walk model.

### A. Algorithm for QMM

Step1. Start

Step2. Read N, lambda, mu, st, s, t

Step3. I = 0

Step4. R = rand(n, 1)

Step5. Iat = -1/lambda * log(r)

Step6. X = zeros(n, 1)

Step5.  Y = zeros(n, 1)

Step6. X(1) = iat(1)

Step7. x(i) =  x(i-1)+ iat(i)

Step9. Y = -1/mu * log(r)

Step10. Y(1) = X(1) + st(1)

Step11. Y(i) = max(x(i) + st(i), y(i-1) + st(i))

Step8. I = i+1

Step12. If i<=n then go to 7

Step13. Plot x, y

Step14. While I < t

Step15. Drawnow

Step16. Pause(p)

Step17. X = x+s*rand(n, 1)

Step18. Y= y+s*rand(n, 1)

Step19. End

International Journal of Managing Information Technology (IJMIT) Vol.4, No.3, August 2012

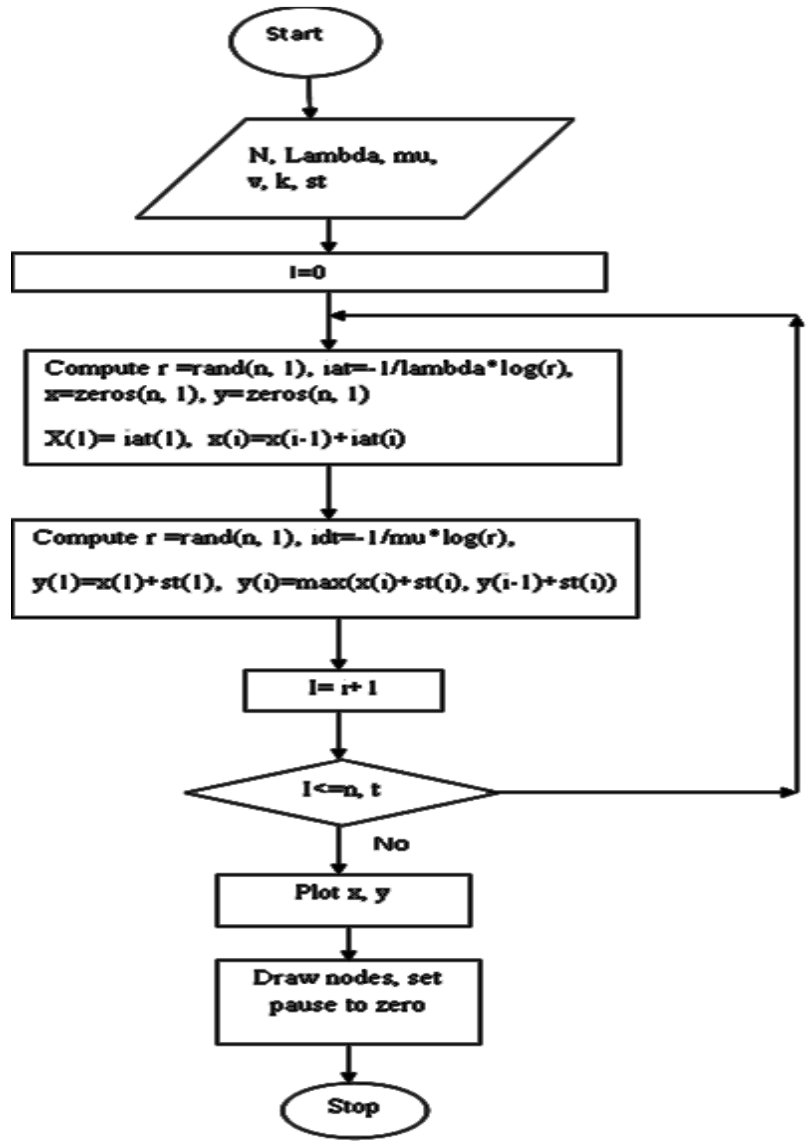

Figure 2: Flowchart For Queueing Mobility Model

## B. Algorithm for Random Walk Model

Step1. Start

Step2. Input n, s, t, i

Step3. X = rand(n, 1)

Step4. Y = rand(n, 1)

Step5. H = plot(X, Y,'.')

Step6. Set axis

Step7. I=0

Step9. While I <t

Step10. Drawnow

Step11. Pause(p)

Step12. X = X+s*rand(n, 1)

Step13. Y= Y+s*rand(n, 1)

Step14. I =i+1

Step15. If I <=t Go to step10

Step16. End

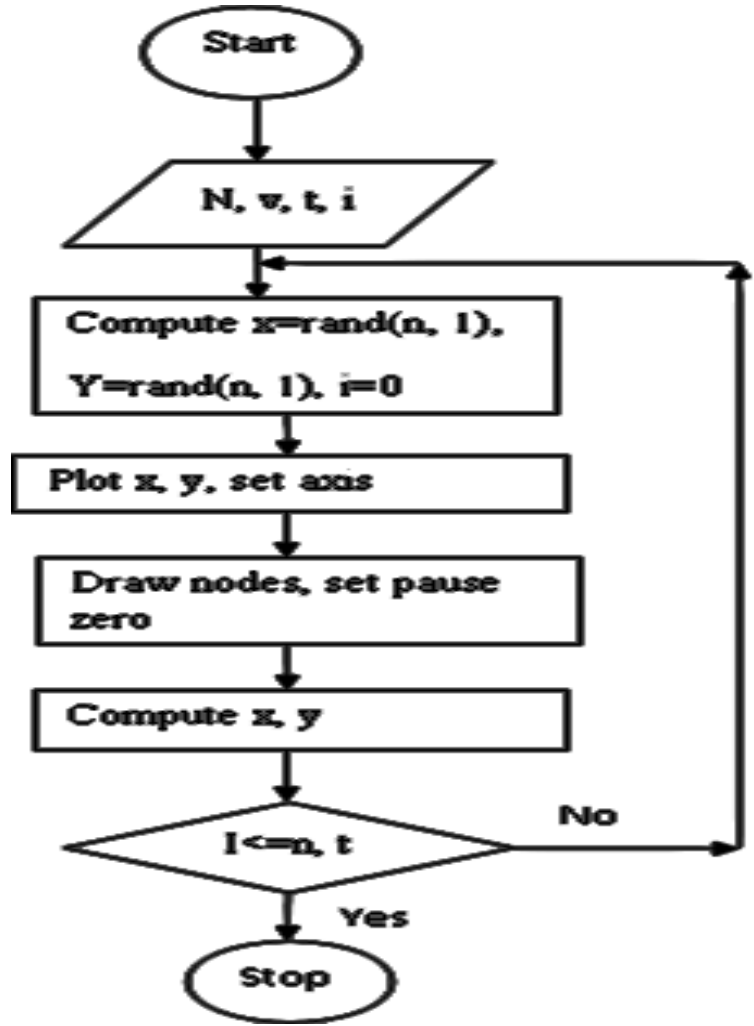

Figure 3: Flowchart For Random Walk Model

### 4.2.3. Simulation of Queuing Mobility Model

The flexibility of MANETs provides that each node can move arbitrarily but in reality, the nodes have a specific behavior which is dependent on the location and the intension of a node. Mobility is one of the main advantages of MANETs where the nodes can move arbitrarily with nearly no restriction. Such movements yield to different node distributions, arrival patterns and to different speeds.

### 4.3. VARYING OF ARRIVAL RATES FOR NODE DISTRIBUTION

The arrival pattern of mobile nodes has an impact on the performance of the network. In this scope, we have decided to analysis the effect of arrival distribution on the MNs population in a given area by using Poisson distribution as in equation 1. In most real-world MANETs, the node population in an area of interest varies with time. In this simulation, it is therefore necessary to investigate the impact of arrivals of MNs on the MANETs mobility.

The simulation area does not change as the arrival rate changes. The different values of arrival rates being considered in this study are shown in Table 2.

Table 2: Varying Arrival Rates

| Scenario | 1 | 2 | 3 | 4 | 5 |
|---|---|---|---|---|---|
| Arrival rates | 0.3 | 0.4 | 0.5 | 0.8 | 0.9 |

During the simulation, nodes were allowed to enter the location from a common source (0 degrees) but not from different sources. The number of MNs that entered the location was assumed to be Poisson distributed with varying arrival rates.

### 4.4. COMPARISON BETWEEN QUEUEING MOBILITY AND RANDOM WALK MOBILITY MODELS

Simulation setup - Starting from an initial node distribution – which represents inhomogeneity, a mobility model was applied to all the nodes with a simulation steps. After the simulation steps elapses a snapshot of the node distribution was observed and analysed. The simulations were based on Queueing Mobility Model (QMM) to study the mobility of MNs from one area to another in a location. The simulations of the models are to be done using MatLab software. The user was allowed to input the number of nodes n, the number of mobility steps to perform for each simulation k and the speed of MNs s. Nodes were initiated from the origin of the simulation area. The simulation area was made up of square area of 300m X 300m. The table 1 shows details of the parameters used for the simulation of both QMM and random walk models.

Table 3:Simulation Parameters

| Parameters | Values |
|---|---|
| Simulation Area | 300m X 300m |
| Number of Nodes | 50, 100, 150, 200, 300 |
| Mobility Steps | 500, 1000, 10000, 20000, 30000, 40000, 50000 |
| Speeds | 0.1m/s to1m/s |

## 5. RESULTS AND DISCUSSION

### 5.1 Effect of Varying Arrival Rates

In Figure 1, the effect of varying nodes' arrival rate is computed using Poisson model. Nodes may arrive at a location either in some regular pattern or in a totally random fashion. The arrival rates have shown to impact on the number of nodes in a particular location, although every location has a limited capacity. A high number of nodes typically translate into a higher average number of neighbours per node, which influences the route availability.

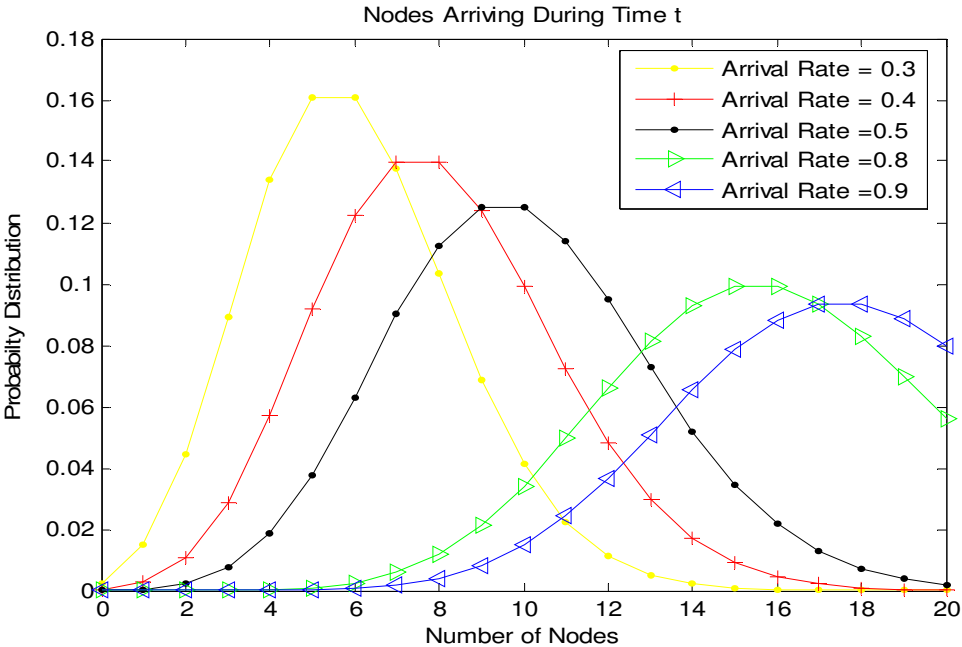

Figure 4: For Twenty Number of Nodes for varying Arrival rates

In reality, the total connection time of a node over a specific interval depends on the nodes encounter rate and the time in each encounter, both of which depend on the relative mobility of nodes.

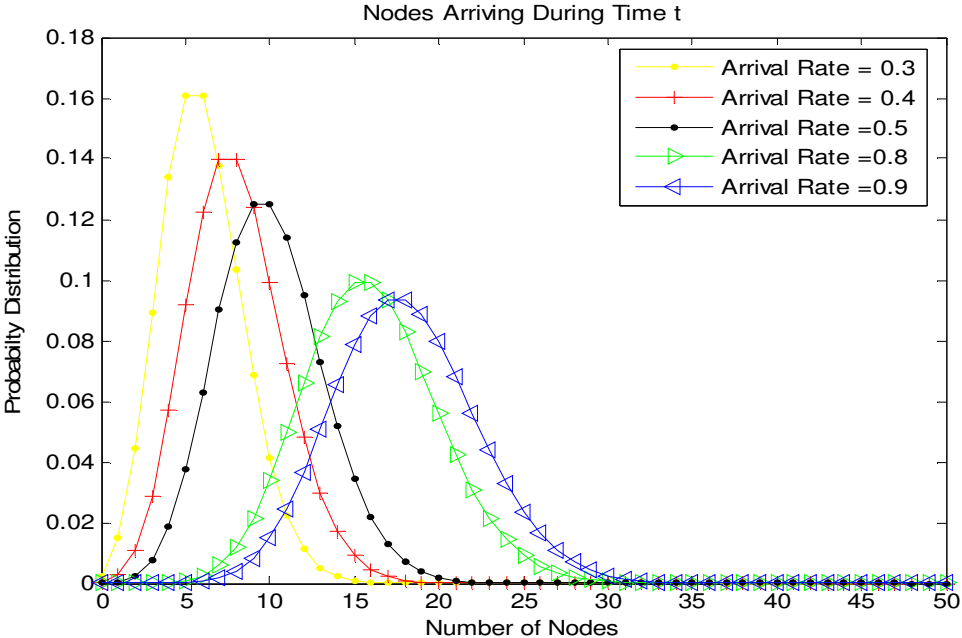

Figure 5: For Fifty Number of Nodes for varying Arrival rates

Although a high node arrivals results in more node encounters, the network would eventually become congested. The impact of this relationship is that nodes can and will be tightly packed (ie. High density) if their arrival rates is high (congestion), but if the arrivals is lower, the nodes must be farther apart (low density). For instance it is clear that there is some congestion for arrivals of MNs, since they have to follow some holding paths.

As the value of arrival rate increases, the shape of the distribution changes dramatically to a more symmetrical ("normal") form and the probability of a larger number of arrivals increases with increasing number of MNs. An interesting observation is that as the arrival rate increases, the properties of the Poisson distribution approach those of the normal distribution as in Figures 1 and 2.

The first arrival processes of nodes give higher contact probabilities at higher arriving rates. This is due to the nodes' contiguity one to another making mobility difficult. In practice, one may record the actual number of arrivals over a period and then compare the frequency of distribution of the observed number of arrival to the Poisson distribution to investigate its approximation of the arrival distribution.

### 5.2 COMPARATIVE STUDY OF MOBILITY MODELS

For the comparative study, we have introduced the notion of a common source where MNs have a same entrance point. We also predefine the number of iterations for the simulation process of the nodes. We compare the QMM against mobility models such as random walk or Brownian mobility models. The program starts with user input window as shown in Figure 6.

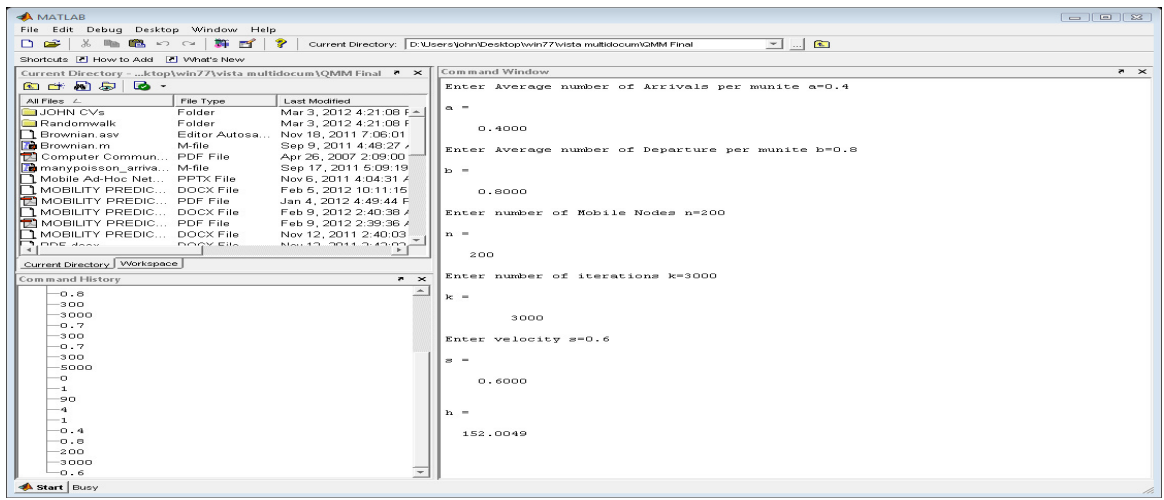

Figure 6: Input command window in Matlab software

The analysis in Figure 7 may indicate that if MNs have been forced to move in a certain order their trajectory therefore is similar to hunts movement. This shows out a typical human behavior when an area is being affected by a natural disaster or unexpected catastrophe. It may recall that a group of hunt always moves in group from a common source point; such a behavior is amendable for self-organized system as in Mobile

Ad'hoc Networks. The radio nodes are assumed to be reconfigurable and acting as router but having a coordinator point, as in MANETs mesh, or cluster topology.

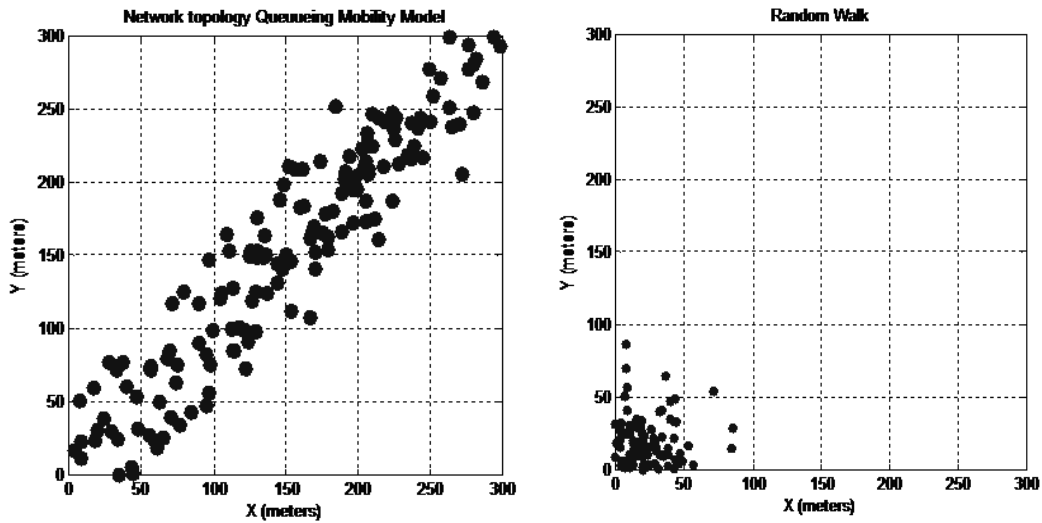

Figure 7: Spatial nodes distribution at the 1000th simulation steps

In Figure 7, the simulation results on the usage of QMM with a chosen common starting point for MNs shows that the MNs span towards to the ground in a given amount of time and take a smaller amount of time to cover an area completely, whereas the obtained results using random walk model with the same parameters such as speed, number of nodes, simulation steps and simulation area do not give the same pattern. This may be explained by the fact that the random walk model, may not give an effective transmission deployment of radio nodes when being used in a case of a natural misfortune alike an earthquake. The radio may not be able to forward the information since it shows a total stacking of radio nodes. Meanwhile for the Poisson model each radio point shows a forwarding movement then, any radio may be viewed as having adequate radius range of transmission. The information reliability is shown to be flexible in the mobility of mobile nodes, with a minimum colliding. It may show a proven for nodes deployment in wireless sensor networks for effective environment monitoring in remote areas.

Emphasis is made on Figures 7 to 9 to illustrate the process of entry to a location from a source, forming line(s) and their departure into various areas within the location.

Each mobility model first places the mobile nodes in their initial locations and defines the way that the nodes move within the network.

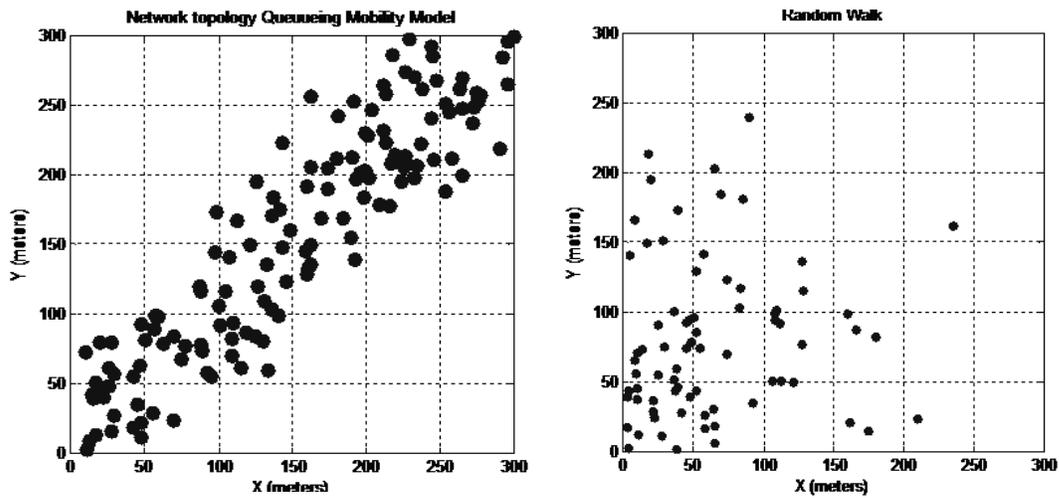

Figure 8: Spatial nodes distribution at the 10000th simulation steps

The QMM may show to have a smaller first entrance time than the random walk model and forms a line or queue from zero degrees to forty-five degrees direction and this leads to a uniform distribution of nodes in steady state. Once nodes have arrived in a location, the nodes will move around the location waiting for the arrival of others. Sometimes MNs arrive in a location waiting for their assign task while others are also joining them resulting to a queue. Hence, the influence of the mobility pattern and/or of the initial distribution on the spatial distribution after a large number of mobility steps on nodes distribution is also observed. The uniformity of the nodes' spatial distribution shows the dependency with the nodes speeds.

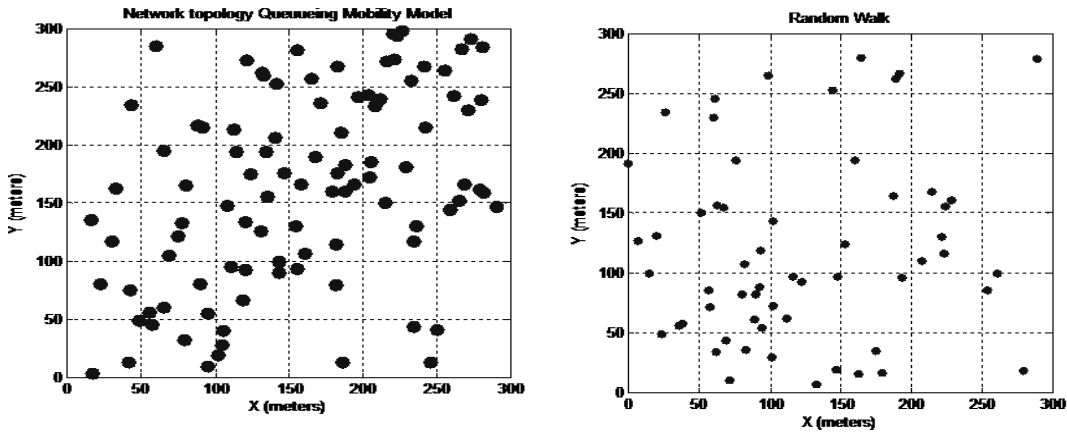

Figure 9: Spatial nodes distribution at the 50000th simulation steps

The observation in Figure 8 to 9 may illustrate that the QMM has "more uniform" nodes distribution compared to the random walk model as the number of simulation steps is higher. In real-world situations, MNs move to a particular destination for specific purposes. More number of nodes may move in groups towards areas of interest or service areas to receive services, sometimes wait or depart or exit immediately. MNs may also select the service areas to visit; there may be other nodes too which are also visiting the same service facility resulting in forming of queues/groups.

The random walk may show a different type of group mobility which may refer to hunts movement when they meet an obstruction or a disturbance a sudden sound they turn around in a scattered manner before proceeding. These may explain why MANETs with a random walk configuration do not find application in military battlefields. The point of entry which was at 0 degree but as they move away from the source to their destinations they begin to disperse just like the QMM. With the QMM nodes are distributed evenly in the simulation area in the long run unlike the random walks which has more nodes at the location closer to the source.

MNS at the front of the queue depart or exit the simulation area as early as possible compared to those at the tail of the queue at the beginning. Some of MNs may return whilst others may move away from the initial group and queue forming subgroups. For a vast simulation period MNs in the group initially dispersed into separate groupings with some of them even moving independently. In the Random Walk model, the mobile nodes were considered moving independently from one another. This kind of mobility model is classified as entity mobility model and however in some scenarios including battlefield communication and museum touring, rescue missions etc., the movement pattern of a mobile node may be influenced by other nodes in the neighborhood.

## 6. CONCLUSION

We have shown the effect of arrival rates on MNs distribution and population in a location. It may claim that as the various arrival rates increased the mean number of MNs may also increased but to a certain limit. It is therefore the indication that every location has a limited capacity. The arrival patterns have shown some impact on the network population, as the arrival rate increases the MNs population also increases to a peak and then decays rapidly to the x-axis.

It may subsequently be admitted that mobility in MANETs is a difficult work and actually. It is an interesting research area that has been growing in recent years. Its difficulty is mainly generated because of the continuous changes in the network topology with time. The topological changes have impact on mobility techniques developed for infrastructure-based networks thus may not be directly applied to mobile adhoc networks. We have investigated through simulation mobility prediction of MNs using the queueing model. It has indicated that the initial MN position may not be uniform (non-homogeneous) at source, however, with time, the distribution of MN positions tend to be homogeneous in the simulation area for QMM. Also with the RWM the distribution of the MN position may be non-homogeneous, that is the trend turn to be dense toward the source and the middle of the simulation area.

Our future work will investigate MNs mobility and its effects on end to end delay and on routing protocol of MANETs. It may be necessary to implement it, by using traces in order to validate the QMM and its reality.

## REFERENCES

[1] J. Boleng, T. Camp, and V. Tolety. "A Suvey of Mobility Models for Ad hoc Network Research", In Wireless Communication and Mobile Computing (WCMC), Vol. 2, No. 5, pages 483 – 502, 2002.

[2] Subir Kumar Sarkar, T. G. Basavaraju, C. Puttamadappa, "Mobility Models for Mobile Ad Hoc Networks", 2007, Page 267 – 277, Auerbach Publications – www.auerbach-publications.com Volume 2, No.5, 2002.

[3] C. Rajabhushanam and A. Kathirvel, "Survey of Wireless MANET Application in Battlefield Operations", (IJACSA) International Journal of Advanced Computer Science and Applications, Vol. 2, No.1, January 2011.

[4] Buttyan L., and Hubaux J. P., "Stimulating cooperation in self-organizing mobile ad hoc networks. Mobile Networks and Applications: Special Issue on Mobile Ad Hoc Networks, 8(5), 2003.

[5] C.P.Agrawal, O.P.Vyas and M.K Tiwari, "Evaluation of Varrying Mobility Models & Network Loads on DSDV Protocol of MANETs", International Journal on Computer Science and Engineering Vol.1 (2), 2009, pp. 40 - 46.

[6] P. N. Pathirana, A. V. Savkin & S. K. Jha. "Mobility modeling and trajectory prediction for cellular networks with mobile base stations". MobiHoc 2003: 213 -221.

[7] Mohd Izuan Mohd Saad and Zuriati Ahmad Zukarnain, "Performance Analysis of Random-Based Mobility Models in MANET Routing Protocol, EuroJournals Publishing, Inc. 2009, ISSN 1450-216X Vol.32 No.4 (2009), pp.444-454 http://www.eurojournals.com/ejsr.htm

[8] Zainab R. Zaidi, Brian L. Mark: "A Distributed Mobility Tracking Scheme for Ad-Hoc Networks Based on an Autoregressive Model". The 6th International Workshop of Distributed Computing, Kolkata, India (2004) 447(458)

[9] Abdullah, Sohail Jabbar, Shaf Alam and Abid Ali Minhas, "Location Prediction for Improvement of Communication Protocols in Wireless Communications: Considerations and Future Directions", Proceedings of the World Congress on Engineering and Computer Science 2011 Vol. II WCECS 2011, October 19-21, 2011, San Francisco, USA

[10] Gunnar Karisson et al., "A Mobility Model for Pedestrian Content Distribution", SIMUTools '09 workshops, March 2-6, 2009, Rome Italy

[11] D.B. Hoang and K.J. Pye. "Introduction to Queueing Theory" – Computer Communication Networks, School of Electronic Engineering, La Trobe University1995 edition (7.54 p.m. 31January, 1996, P91-104)

[12] Ivo Adan and Jacques Resing. "Queueing Theory", Department of Mathematics and Computing Science, Eindhoven University of Technology, The Netherlands, February 14, 2001, pp. 29 – 32)

[13] Anreas Willig, "A Short Introduction to Queueing Theory", Technical University Berlin, Telecommunication Networks Group, Berlin, July 21, 1999, page 3 - 4.

[14] M.E. Kohl, N.M. Steiger, F.B Armsstrong and J.A. Joines, "An efficient Performance Extrapolation For Queueing Models in Transient Analysis", Proceedings of the 2005 Winter Simulation Conference.

[15] "Mathworks, Inc, developer and distributor of technical computing software matlab." http://www.mathworks.com/products/matlab.

## Authors

Short Biography

**John Tengviel**

He is a holder of a BSc. Computer Science from University of Science and Technology (KNUST) in 2001 and currently candidate of MSc. Telecommunication Engineering from College of Engineering at the same university. He is a senior instructor with the Department of Computer Science at Sunyani Polytechnic. His research interests include Mobile Ad hoc Networks and Mobility modeling in MANETs.

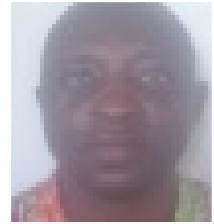

**K. A. Dotche**

He is a holder of a BSc. Electrical Eng. from University of Lome and MSc. Telecom. Eng. from College of Engineering at Kwame Nkrumah University of Science and Technology (KNUST); respectively 2004 and MSc 2010. He is currently a Ph. D. research candidate with the Department of Telecommunications Engineering, at KNUST. His research interests include Energy efficiency in wireless sensor networks, antennas and E-M propagation in cellular layered networks.

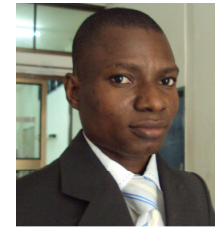

**Nana (Dr.) Kwasi Diawuo** is a senior lecturer of the Department of Computer Engineering at Kwame Nkrumah University of Science and Technology (KNUST), Kumasi, Ghana. He earned a BSc. (Electrical/ Electronic Engineering) from KNUST, M.Sc., Ph.D, MGhIE. He is a member of the Institution of Electrical and Electronic Engineers (IEEE) and Computer Society (of IEEE).